# Electronic structure of GaSb/AlGaSb quantum dots formed by filling droplet-etched nanoholes

L. Leguay,[1,*] A. Chellu,[2,*,†] J. Hilska,[2] E. Luna,[3] A. Schliwa,[1] M. Guina[2] and T. Hakkarainen[2]

[1] *Institut für festkörperphysik, Fakultät II, Technische Universität Berlin, Hardenbergstraße 36, 10623, Berlin, Germany*

[2] *Optoelectronics Research Center, Physics Unit, Korkeakoulunkatu 3, 33720, Tampere, Finland*

[3] *Paul-Drude-Institut für Festkörperelektronik, Leibniz-Institut im Forschungsverbund Berlin e.V., Hausvogteiplatz 5-7, 10117 Berlin, Germany*

\* LL and AC contributed equally to this work

† Corresponding author: abhiroop.chellu@tuni.fi

Epitaxially-grown semiconductor quantum dots (QDs) provide an attractive platform for the development of deterministic sources of high-quality quantum states of light. Such non-classical light sources are essential for quantum information processing and quantum communication. QDs emitting in the telecom wavelengths are especially important for ensuring compatibility with optical fiber systems required to implement quantum communication networks. To this end, GaSb QDs fabricated by filling local-droplet etched nanoholes are emerging as a viable approach, yet the electronic properties of such nanostructures have not been studied in detail. In this article, an insight into the electronic structure and carrier dynamics in GaSb/AlGaSb QDs is provided through a systematic experimental analysis of their temperature-dependent photoluminescence behavior. A steady-state rate equation model is used to reveal the relevant energy barriers for thermally activated carrier capture and escape processes. Furthermore, results of detailed theoretical simulations of quantum-confined energy states using the multi-band k·p model and the effective mass method are presented. The purpose of the simulations is to reveal the direct and indirect energy states, carrier wavefunctions, and allowed optical transitions for GaSb QDs with different physical dimensions.

## I. INTRODUCTION

Semiconductor quantum dots (QDs) are excellent deterministic sources of single photons and entangled photon pairs, which are instrumental for the deployment of photonic quantum information technologies. Non-classical light emission in the technologically important third telecom window is highly desirable for quantum communication protocols exploiting the global absorption minimum in optical fibers and key atmospheric transmission windows for free-space operation [1]. Moreover, integrating QDs emitting in the telecom wavelengths with the silicon photonics platform offers a scalable approach to realizing quantum photonic integrated circuits with low waveguide propagation losses [2,3]. Strain-free epitaxial GaAs/AlGaAs QDs fabricated by filling nanoholes formed via local droplet etching (LDE) have already demonstrated state-of-the-art metrics for single-photon indistinguishability and entanglement fidelity in the 680 – 800 nm spectral range [4–7]. While considerable effort has been made to develop epitaxial QDs emitting in the telecom O- and C-bands using various growth techniques and material compositions [8–10], their general performance in terms of non-classical optical properties still lags behind the state-of-the-art demonstrations by InGaAs/GaAs QDs and GaAs/AlGaAs QDs emitting at shorter wavelengths [11–13].

Strain-free GaSb/AlGaSb QDs grown by in-filling LDE-nanoholes have been recently proposed as deterministic sources of non-classical light emission at telecom wavelengths. These QDs exhibit narrow exciton lines and single-photon emission in the telecom S-band with state-of-the-art homogeneity for a self-assembled growth process [14–16], making them an excellent candidate for quantum communication. From a device perspective, QD-based single-photon sources greatly benefit from employing resonant [12,17] or quasi-resonant [9,18–20] excitation schemes which, however, require precise knowledge about the available energy states in the QDs. Moreover, advancing this technology towards application requires an understanding of the relation between the growth parameters of the QDs and intricate knowledge of the electronic structure. This enables to tune the available energy states, and is especially important in the case of GaSb QDs which exhibit the unique feature of QD-size-dependent indirect-direct bandgap crossover [14]. Therefore, a detailed understanding of the energy structure including radiative transition energies, barrier energies involved in carrier movement, quantum confinement effects and allowed optical transitions is necessary for moving towards practical integration of this emerging QD material in quantum photonic devices. In this paper, we combine theoretical modeling and temperature-dependent photoluminescence (PL) experiments to gain insight into the carrier dynamics and electronic structure of GaSb/AlGaSb QDs.

## II. METHODS

### A. Epitaxy and sample structure

The samples investigated in this work were grown by molecular beam epitaxy on n-GaSb(100) substrates. We fabricated GaSb QDs via the LDE [21] technique where nanoholes are formed on an AlGaSb matrix using Al droplets and are subsequently filled by GaSb. The LDE process involved depositing 1.5 monolayers (ML) of Al at 395 °C to form Al droplets that subsequently etch the underlying AlGaSb layer for 180 s under a low Sb-flux of 0.067 ML/s to form nanoholes. More details related to LDE in the GaSb/AlGaSb system and QDs formed by this technique are available in our previous publications [14,15]. Atomic force microscopy (AFM) imaging of as-grown nanoholes on an AlGaSb surface helped in assessing their physical dimensions, which directly influence the optical emission properties of the QDs. Additionally, the analysis of the AFM images of the as-grown nanoholes revealed a density of $2.6 \times 10^7$ cm$^{-2}$, average nanohole diameter of ~50 nm and average nanohole depth of 12 nm (Fig. 1(a)). The nanohole depth is of particular relevance here because it directly determines the physical height of the GaSb QDs formed by nanohole-filling.

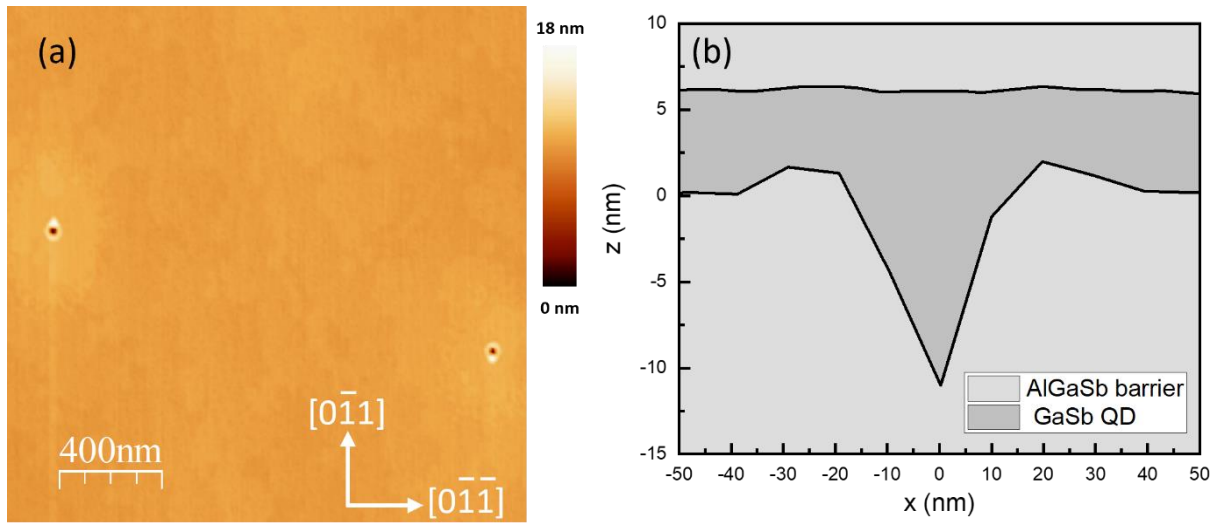

FIG. 1. (a) 2x2 µm AFM image of as-grown nanoholes in AlGaSb. (b) AFM profile before and after filling a nanohole with GaSb. The line profiles are offset by 6.1 nm to illustrate the estimated geometry of the QDs in QD$_{6.1}$.

Two samples named $QD_{5.7}$ and $QD_{6.1}$ containing nanoholes filled with 5.7 nm and 6.1 nm of GaSb respectively were used in PL experiments. AFM imaging carried out after the nanohole filling stage in a separate sample revealed a planarized top surface, which is conclusive evidence that all the nanoholes are completely filled with GaSb (Fig. 1(b)). The samples used for PL experiments contain GaSb QDs sandwiched between AlGaSb barriers and AlAsSb claddings as shown in the chemically-sensitive $\mathbf{g}_{002}$ dark-field (DF) transmission electron microscopy (TEM) image in Fig. 2. Cross-sectional TEM specimens were prepared in the [110] and [1-10] projections using mechanical thinning, followed by Ar-ion milling. In order to minimize the sputtering damage, the Ar-ion energy was reduced to 2 – 0.5 keV. The samples were investigated on a JEOL 3010 microscope and on a (S)TEM JEOL 2100F microscope both operating at 200 kV. The samples are of high structural perfection and no extended defects are detected. Chemically sensitive $\mathbf{g}_{002}$ DF-TEM is a powerful and direct method to determine elemental distribution in III-V semiconductors with zinc-blende structure [22,23]. When 002 imaging conditions are properly set up (the specimen is tilted 8–10º from the <110> zone axis towards the [100] pole while keeping the interface edge-on), the contrast directly reflects the chemical composition of the alloy [23]. Under these imaging conditions, AlGaSb layers show a brighter contrast compared to GaSb, with the AlGaSb contrast getting brighter as the Al content increases. As observed in Fig. 2(a) and more specifically in the zoom-in micrograph in Fig. 2(b), filling with GaSb planarizes the growing surface (note that no QDs are detected in these images) giving rise to a GaSb quantum well (QW). The average QW thickness for sample $QD_{5.7}$ is 5.7 ± 0.1 nm, estimated from the analysis of images obtained along both [110] and [1-10] zone axes. The error bar refers to the standard deviation of the data. Furthermore, $\mathbf{g}_{002}$ DF-TEM imaging clearly reveals the presence of a very bright band of ~1.4 nm width below the QW. An estimation of the Al content in the thin layer based on the analysis of the $\mathbf{g}_{002}$ diffracted intensity ($I_{002}$) reveals a very high Al content [Al] > 70%. The narrow band is indeed an Al-rich pre-layer which has been unintentionally formed during the Al-wetting of the surface prior to the droplet formation.

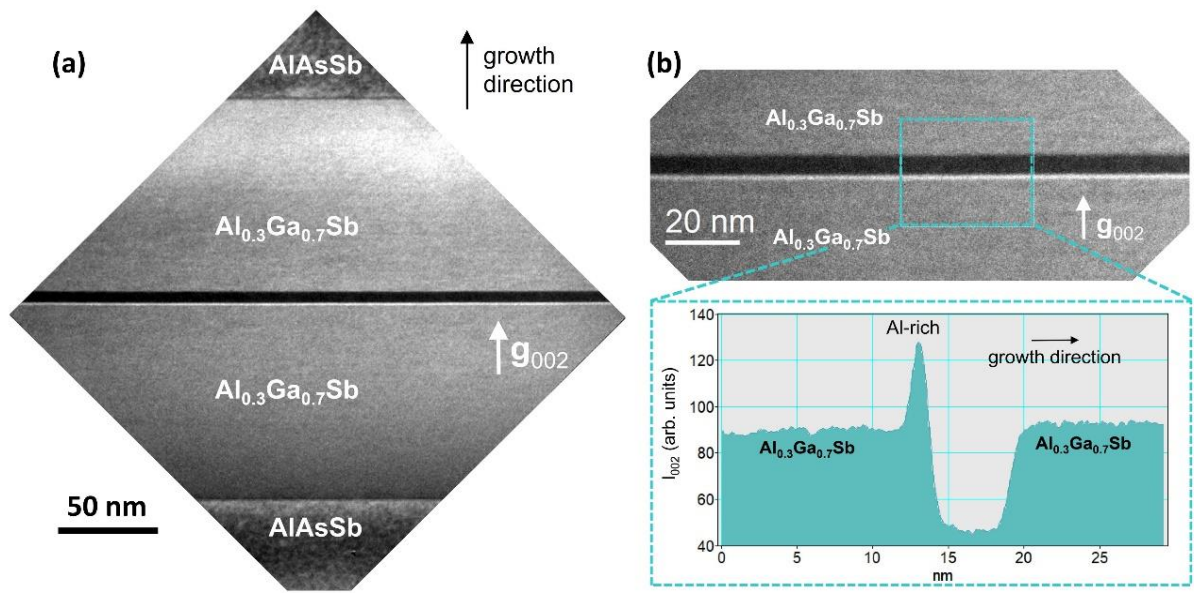

FIG. 2. (a) Chemically-sensitive $\mathbf{g}_{002}$ DF-TEM overview image of the layer structure in sample $QD_{5.7}$ and (b) zoom-in micrograph of the QW region together with a representative intensity line scan obtained from the area marked in the image. Under $\mathbf{g}_{002}$ DF-TEM imaging conditions, the thin Al-rich pre-layer that is formed at the lower GaSb/AlGaSb interface during the droplet formation step of the LDE process appears as a bright band below the QW.

## B. Photoluminescence experiments

Temperature-dependent PL (TDPL) measurements were carried out with the sample(s) placed inside a closed-cycle He cryostat. A 532 nm laser focused down to ~1 mm in diameter on the sample was used to non-resonantly excite an ensemble of QDs while the temperature was varied from 11 K – 150 K. TDPL measurements were made with two different excitation laser power densities (0.025 W cm$^{-2}$ and 0.13 W cm$^{-2}$) to clearly identify the emission from ground and excited states of the QDs. The PL signal was analyzed by passing it through a 500 mm monochromator and detected with a Peltier-cooled InGaAs photodetector.

## C. Quantum mechanical model

The energy levels and wavefunctions of bound electron and hole states are calculated using the eight-band k·p model in real space, employing non-periodic boundary conditions. The theory was initially developed for the description of electronic states in bulk materials [24–26]. For the use in heterostructures, the envelope-function version of the model has been developed and applied to QWs [27], quantum wires [28], and QDs [29–32]. Details of the principles of our implementation for zinc-blende heterostructures are outlined in [32]. This model enables us to treat QDs of arbitrary shape and material composition, including the effect of strain, piezoelectricity, valence band (VB) mixing, and conduction-valence band interaction. Due to the limited number of Bloch functions used for the wavefunction expansion, the results of the eight-band k·p model are restricted to the vicinity of the center of the Brillouin zone. Off-center states related to L- and X-points are calculated using the effective mass model.

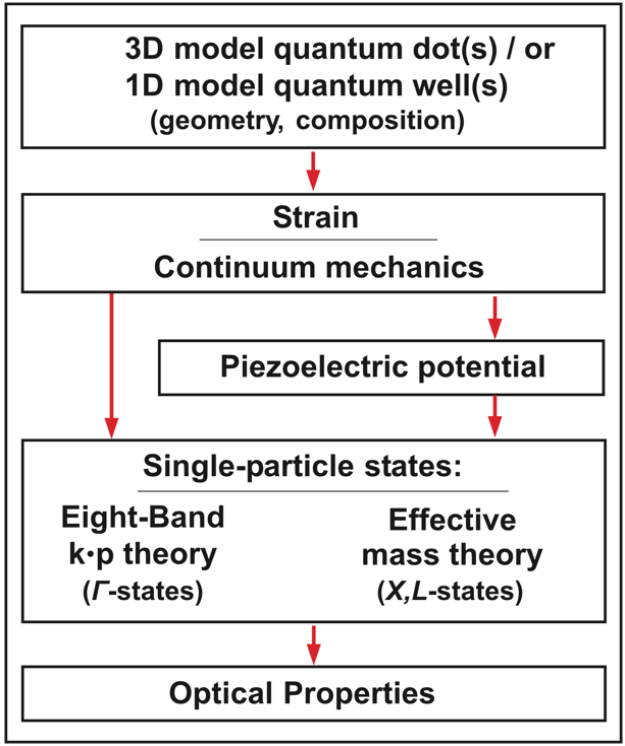

FIG. 3. Schematic of the modeling procedure applied in this work.

Fig. 3 shows a schematic of the modeling procedure followed in this work. The process starts by implementing the 3D structure of the QD model, including its size, shape, and chemical composition, followed by a calculation of the strain and piezoelectricity. The resulting strain and polarization fields are used in the eight-band k·p Hamiltonian. Solving the Schrödinger equation allows us to determine the electron and hole single-particle states. Finally, optical properties such as absorption spectra, capture cross-sections and lifetimes can be calculated.

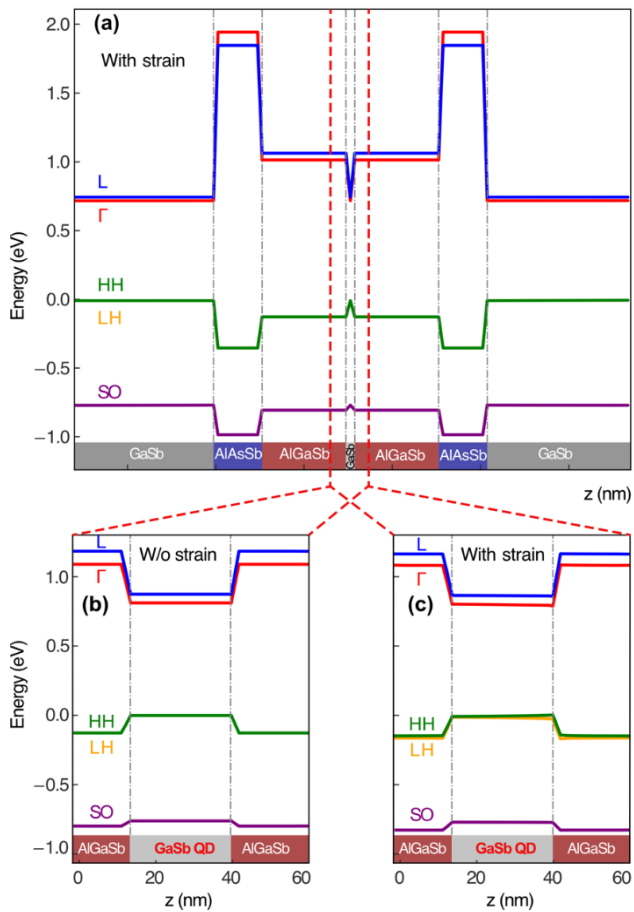

FIG. 4. (a) Local band-edges for the whole structure, including the effect of strain. (b) and (c) Zoomed-in view into the quantum region with the GaSb QD's band-edge structure calculated without and including strain respectively.

All the simulations were performed using the nextnano++ software developed by nextnano GmbH [33]. This software is a Schrödinger-Poisson-current solver specifically designed for semiconductor calculations and is capable of handling both 1D and 3D nanostructures. The simulations were performed using both the effective mass method (L- and X-states in 3D) and the 8-band k·p model (Γ-states in 1D and 3D). The effective mass method was employed for studying the bandgap crossover. The single-band effective mass Schrödinger equation was solved for each selected conduction band (CB), with the energies of the Γ- and L-bands being calculated and compared to determine whether the bandgap was direct or indirect. For the study of dipole transitions, the 8-band k·p Schrödinger equation was solved for the Γ-CB, as well as the heavy, light, and split-off hole VBs. Ten electron states and ten hole states were simulated, and the inter-band and intra-band matrix elements, transition energies, and oscillator strengths were calculated.

For an overview of the local band-edge variation of the whole device, as shown in Fig. 4(a), a 1D model was created by stacking various materials, including a GaSb substrate, Al$_{0.3}$Ga$_{0.7}$Sb cladding, and a GaSb QW at the center, which forms the quantum region. Next, the band-edges of the quantum region with the GaSb QD in the center are treated as 3D structure, and their corresponding band-edges are shown in Fig. 4(b) in absence and in Fig. 4(c) in presence of strain. As the cladding layer composition is chosen in a way that the lattice mismatch is minimal, and the QD itself is made of the matrix material, almost no strain, hence, piezoelectricity is present inside the QD, as can be seen by comparing Figs. 4(b) and 4(c).

In the 3D model, a GaSb QD with an inverted cone shape was added beneath the QW (see Fig. 5). The key dimensions of the structure include the depth and diameter of the QD and the thickness of the QW.

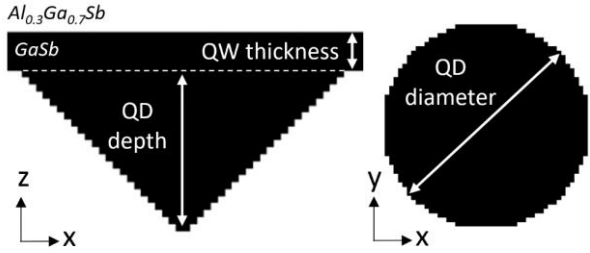

Fig. 5. Schematic of the 3D modeled structure.

## III. RESULTS AND DISCUSSION

### A. Temperature-dependent photoluminescence

We first analyze the TDPL spectra collected from QD$_{5.7}$ and QD$_{6.1}$ shown in Fig. 6 and identify the origin of the various peaks. Figs. 6(a) and 6(c) show spectra measured from the two samples with an excitation power density of 0.025 W cm$^{-2}$. The peak starting from a center wavelength of ~1320 nm (at low temperature) is the ground state emission from the GaSb QW that covers the surface planar areas of the bottom AlGaSb layer and is labelled as QW in the plots. Next, the peak starting with a low-temperature center wavelength of ~1470 nm is the ground state emission from the GaSb QDs and is labelled as GS in the plots. The peak with a slightly longer wavelength than QW and clearly evident in the temperature range 11 K – 60 K is presumed to be emission from carriers in a localized state outside the nanohole (labelled as P). Power-dependent PL analysis of this emission peak in our previous work has revealed behavior akin to the ground state of a localized system, i.e. with a linear power-dependency [14], which in this case could be the carrier localization in the QW states around AlGaSb nanoring surrounding the nanohole. Moving on to the spectra in Figs. 6(b) and 6(d) which were collected with a higher excitation power density of 0.13 W cm$^{-2}$, we see the emergence of an additional peak between P and GS. GS emission intensity is seen to saturate at this high excitation power density as a result of carriers fully occupying the ground state and starting to populate the next available energy level in the QDs. Thus, the peak labelled as ES arises from the radiative recombination of electrons and holes in the first excited conduction and valence states of the QDs, respectively. Consequently, ES peak exhibits a superlinear power-dependency [14].

A cursory look at the TDPL spectra from the two samples immediately reveals three things: first, we see that all the emission peaks from QD$_{6.1}$ are at slightly longer wavelengths compared to the emission peaks from QD$_{5.7}$. This suggests a stronger quantum confinement effect for QDs with 5.7 nm filling resulting in discrete states at slightly higher energies than for the QDs with 6.1 nm filling. Next, we note that the center wavelength of all peaks in the TDPL spectra monotonously red-shift as the temperature increases. This behavior is expected due to bandgap shrinkage as the temperature increases and has been well-documented in a variety of bulk semiconductor systems [34,35]. However, it has been shown that the rate of bandgap shrinkage for low-dimensional systems like QWs, nanowires and QDs deviates from the rate seen in bulk structures especially in the low-temperature regime [36–40]. In section III B, we use semi-empirical analytical models to study this non-linear red-shift of the QD ground state emission and determine quantitative values for material parameters that cause this behavior. The final aspect of the TDPL spectra that we clearly notice is the temperature-dependency of the emission peak(s) intensity which we will discuss in detail in section III C.

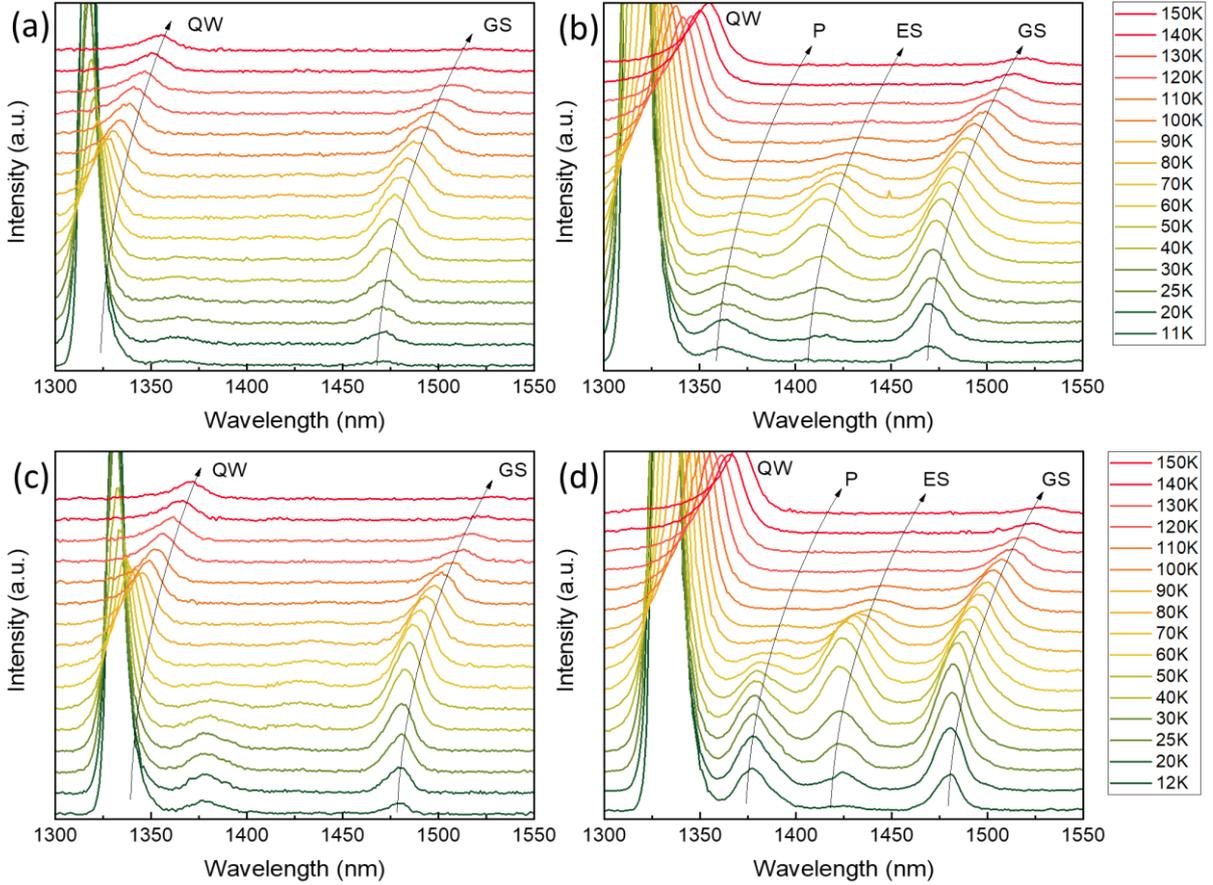

FIG. 6. PL spectra from (a) $QD_{5.7}$ and (c) $QD_{6.1}$ for temperatures ranging from ~11 K to 150 K collected using an excitation power density of 0.025 W cm$^{-2}$. (b) and (d) show TDPL spectra collected using a higher excitation power density of 0.13 W cm$^{-2}$.

### B. Temperature-dependency of the peak energy

The peak energy of the GS emission ($E_{GS}$) is extracted from the TDPL spectra of $QD_{5.7}$ and $QD_{6.1}$ (in Figs. 6(a) and 6(c)) and is plotted as a function of temperature in Fig. 7. We see that $E_{GS}$ rises linearly as the temperature is reduced from 150 K till 80 K, below which the E(T) behavior becomes non-linear.

As the first step in analyzing the experimentally observed E(T) behavior, we use the well-known semi-empirical Varshni relation

$$E(T) = E(0) - \frac{\alpha T^2}{\beta + T} \quad [41]. \tag{1}$$

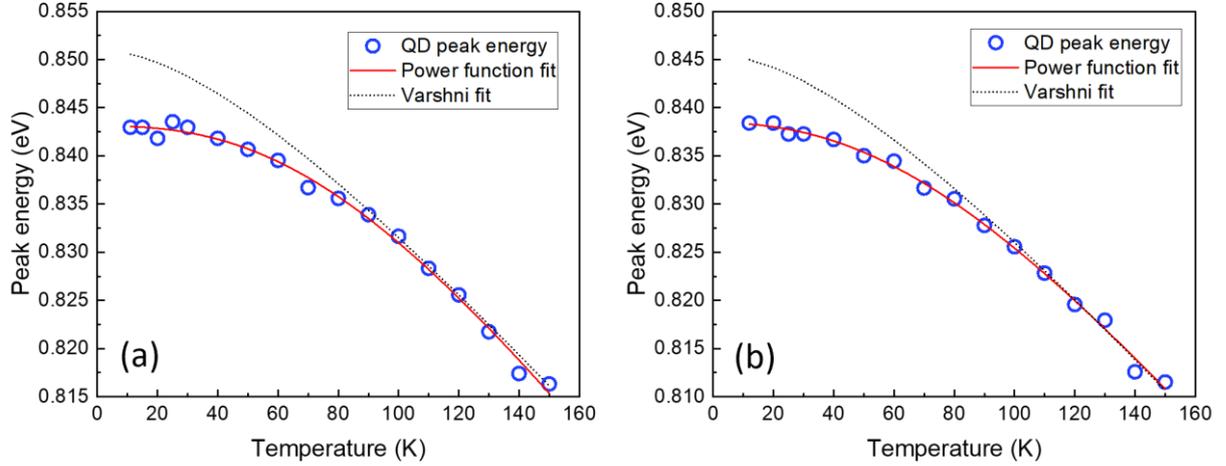

FIG. 7. E(T) data plotted for GS peak from TDPL spectra of (a) QD$_{5.7}$ and (b) QD$_{6.1}$. The experimental E(T) data are fitted using Varshni's model Eq. (1) and the power function model Eq. (3).

Fitting the experimental E(T) data using Eq. (1) fails at low temperatures with any physically reasonable combination of α and β. The best fit for the intermediate temperature range is obtained using bulk values for α and β (Fig. 7) and using E(0) as a free parameter. The shortcomings of the Varshni model, especially for fitting E(T) behavior in the low-temperature regime, have already been reported [42,43].

Phonon-dispersion based models such as the two-oscillator model and the power function model have been used to successfully describe the E(T) behavior in InGaAs QDs over a wide temperature range [37]. According to dispersion-based semi-empirical models, two physical mechanisms namely, electron-phonon interactions and thermal lattice expansion, contribute to bandgap reduction in bulk semiconductors as well as in low-dimensional nanostructures like QWs, nanowires and QDs. The cumulative contribution of both mechanisms is represented by

$$E(T) = E(0) - \frac{\alpha}{k_B} \int \frac{\varepsilon w(\varepsilon) d\varepsilon}{\exp\left(\frac{\varepsilon}{k_B T}\right) - 1} \quad [42]. \tag{2}$$

Here, $\varepsilon = \hbar\omega$ is the energy of the individual lattice oscillators and is proportional to the average occupation number of phonons $\bar{n}=[\exp(\varepsilon/k_B T)-1]^{-1}$. The constant $\alpha=-dE(T)/dT|_{T\to\infty}$ corresponds to the limiting slope of the E(T) dependence in the high temperature limit and $w(\varepsilon) \geq 0$ is the normalized weighting function of the phonon oscillators. Assuming different values for $w(\varepsilon)$ leads to different analytical models, one of which is the power function model where an ansatz for the weighting function in the form of $w_\nu(\varepsilon)=\nu(\varepsilon/\varepsilon_c)^{\nu-1}/\varepsilon_c$; $\nu=0$ for $\varepsilon>\varepsilon_c$ gives

$$E(T) = E(0) - \frac{\alpha\theta}{2}\left\{\left[1 + \sum_{n=1}^{3} a_n(\nu)\left(\frac{2T}{\theta}\right)^{n+\nu} + \left(\frac{2T}{\theta}\right)^{5+\nu}\right]^{\frac{1}{(5+\nu)}} - 1\right\} \tag{3}$$

with the expansion coefficients $a_n(\nu)$

$$a_1(v) = \frac{5+v}{6}\left(\frac{\pi}{2}\right)^{2+(v-1)^2/2} \, , \, a_2(v) = \frac{1-v}{2} \, , \, a_3(v) = \frac{(5+v)(1+v)^2}{3v(2+v)} \quad (4)$$

The constants in Eq. (3) are the average phonon temperature $\Theta=\langle\varepsilon\rangle/k_B=v\varepsilon_c/[k_B(1+v)]$ and cut-off energy $\varepsilon_c=k_B\Theta(1+v)/v$. The material-specific dispersion coefficient ($\Delta$) is defined using the curve shape parameter ($v$) as $\Delta=1/\sqrt{v(2+v)}$ which quantifies the different phonon dispersion regimes namely, moderate dispersion regime for $0<\Delta<3^{-1/2}$ and large dispersion regime $3^{-1/2}<\Delta<\infty$. A parameter set consisting of the dispersion coefficient ($\Delta$), the effective average phonon temperature ($\Theta$), the limiting slope ($\alpha$) and the material bandgap at 0 K (E(0)) allows us to unambiguously determine the E(T) dependence in a wide temperature range. Fitting the power function model shown in Eq. (3) to the experimental E(T) data sets in Fig. 7 converges very well even at lower temperatures. The resulting fitting-parameter set is shown in Table I. The values obtained for $\Theta$ and $\Delta$ are in reasonably good agreement with the data available in literature for antimonide bulk materials and in the same range as what has been reported for InGaAs/GaAs QDs [34]. It should be noted also that the fitting results for $\Delta$ fall within the moderate dispersion regime, where the Varshni model fails to reproduce the experimentally observed low-temperature behavior [44].

TABLE I. Parameter set obtained from fitting the power function model [Eq. (2)] to the experimental E(T) data in Figs. 7(a) and 7(b).

| Parameter | $QD_{5.7}$ | Error | $QD_{6.1}$ | Error |
|---|---|---|---|---|
| $\alpha$ (meV K$^{-1}$) | 4.57E-4 | ± 1.2E-04 | 3.99E-04 | ± 9.4E-05 |
| $v$ | 1.4912 | ± 0.61 | 1.1977 | ± 0.51 |
| $\Theta$ (K) | 260 | ± 87 | 230 | ± 82 |
| E(0) (eV) | 0.8431 | ± 4.5E-04 | 0.8384 | ± 5.3E-04 |
| $\varepsilon_c$ (meV) | 37.43 | - | 36.37 | - |
| $\Delta$ | 0.54 | - | 0.56 | - |

### C. Temperature-dependency of intensity

Integrated PL intensity of QW and GS peaks from TDPL spectra (in Figs. 6(a) and 6(c)) is plotted as a function of temperature in Fig. 8. The PL intensity from the QW shows typical temperature dependency as it monotonously decreases with increase in temperature. On the other hand, the GS peak intensity first increases as the temperature is raised from 12 K to 50 K and then decreases rapidly at higher temperatures. This behavior suggests the presence of a low-temperature, thermally activated barrier for carriers to move from the QW to the QD energy levels. We introduce a rate equation model to quantify the carrier transfer mechanisms between the QD and its surroundings. We assume a three-level system containing the top AlGaSb barrier, GaSb QW and GaSb QDs. The AlGaSb layer acts as an infinite reservoir of photo-excited carriers. The carriers are generated in the QW from AlGaSb at a rate $g_{QW}$ and the total number of available carriers is $n_{QW}$. Some of the carriers in the QW recombine radiatively at a rate $r_{QW}$. The remaining carriers can either thermalize to a nearby QD or escape to the surrounding bulk layers at a rate $\alpha_{QW}=e_{QW}\exp(-E_{QW}/k_BT)$. However, the movement of carriers from QW to QDs is mediated by a thermally activated barrier state that considerably traps carriers at low temperatures. Carriers are generated in QDs directly from the AlGaSb reservoir at a rate $g_{QD}$ and by thermalization from the QW via the barrier state at a rate $\alpha_b=e_b\exp(-E_b/k_BT)$. The carriers in the QDs then either thermally escape to the surrounding bulk at a rate $\alpha_{QD}=e_{QD}\ exp(-E_{QD}/k_BT)$ or recombine radiatively at a rate $r_{QD}$. A schematic of the carrier movement described above is shown in Fig. 9.

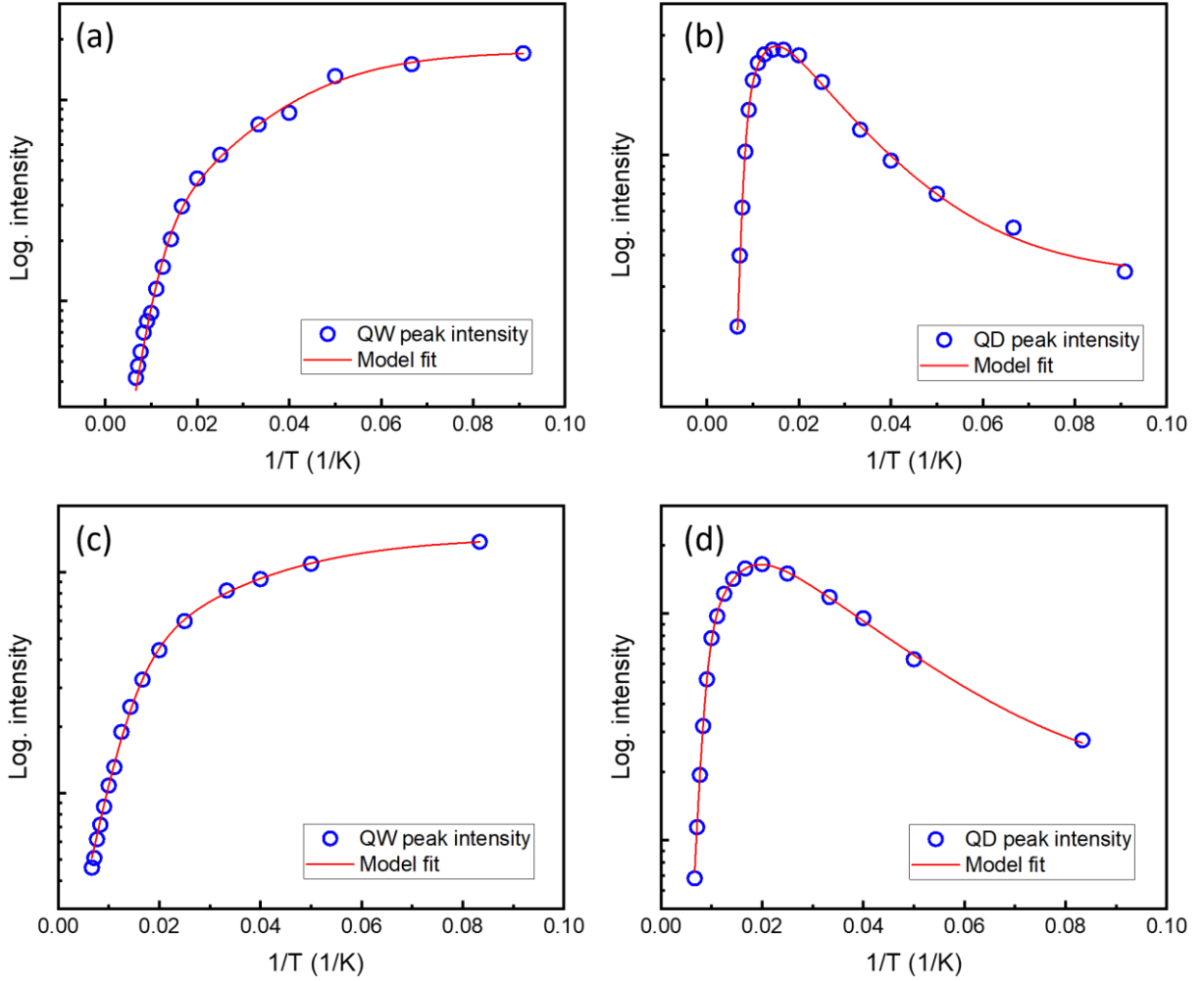

FIG. 8. Integrated PL intensity plotted as a function of temperature for (a) QW peak in QD$_{5.7}$, (b) GS peak in QD$_{5.7}$, (c) QW peak in QD$_{6.1}$, (d) GS peak in QD$_{6.1}$. The integrated intensity is calculated from TDPL spectra collected with an excitation power density of 0.025 W cm$^{-2}$. The red line in the plots represent the steady state-solutions of the rate equations for QW and QDs Eqs. (6) and (7).

The rate equations for the 3-level system describing the movement of carriers between the QW and QDs are:

$$\frac{dn_{QW}}{dt} = g_{QW} - \alpha_{QW}n_{QW} - \alpha_b n_{QW} - r_{QW}n_{QW} \tag{4}$$

$$\frac{dn_{QD}}{dt} = g_{QD} + \alpha_b n_{QW} - \alpha_{QD}n_{QD} - r_{QD}n_{QD} \tag{5}$$

The temperature dependency of PL intensity from QW and QDs is obtained from the steady-state solution of the coupled rate equations.

$$I_{QW}(T) = \frac{g_{QW}}{r_{QW} + \alpha_{QW} + \alpha_b} \tag{6}$$

$$I_{QD}(T) = \frac{[g_{QD}(r_{QW} + \alpha_{QW} + \alpha_b)] + \alpha_b g_{QW}}{(r_{QW} + \alpha_{QW} + \alpha_b)(r_{QD} + \alpha_{QD})} \qquad (7)$$

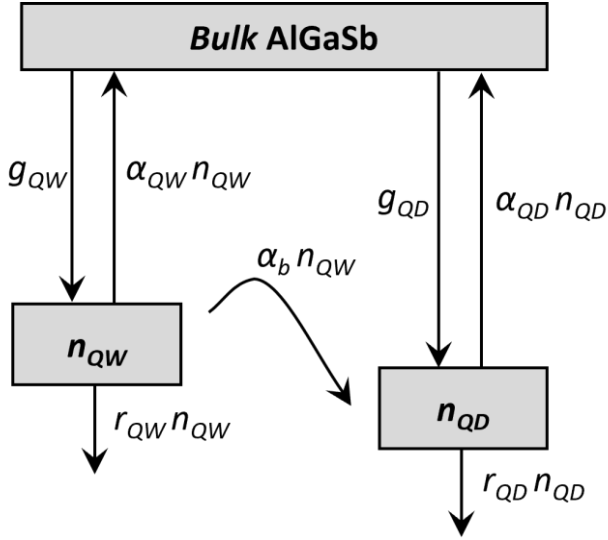

FIG. 9. Schematic representation of the rate equation model.

The temperature-dependent QW and GS peak data in Fig. 8 are fitted with the obtained steady-state rate equation solutions shown in Eqs. (6) and (7). The fitting parameters $E_{QW}$, $E_b$ and $E_{QD}$ represent the activation energies for the carrier escape mechanisms from the QW, barrier and QDs respectively and are summarized in Table II. The fitting parameter $E_{QD}$, corresponding to carrier escape from QDs, has an activation energy in the same range as the energy separation between QW and GS peaks ($\Delta E_{QW-QD}$). Therefore, it is reasonable to consider that the GS emission intensity at higher temperatures is likely limited by excitonic escape to the QW [36,38,45]. The barrier $E_b$ has an activation energy of ~5 meV for injecting carriers from the QW to the QDs and limits carrier injection in the low-temperature range as shown by the rise in intensity of GS peak up to ~50 K (Fig. 6). The presence of such an energy barrier agrees well with the time-resolved PL response of the GaSb QDs collected with above-bandgap excitation, which show a relatively slow radiative decay rate that is dominated by the trapping of carriers during their movement from QW to QDs. Additionally, the second-order autocorrelation function obtained from these QDs exhibits signatures of a carrier reservoir effect [16]. Therefore, from a practical application point of view, a more efficient way to inject carriers would be to excite them to the QDs resonantly either to the GS or ES transition which has been the preferable method for state-of-the-art single- and entangled-photon sources [5,46,47]. An alternative to resonant excitation schemes would be to efficiently transfer non-resonantly excited carriers to the QDs directly from the AlGaSb barrier (thus bypassing the QW-barrier path). This can be realized, for instance, by exploiting the Al-rich layer below the QW (see Fig. 2) to function as a barrier for blocking carriers injected in the AlGaSb layer from transferring to the QW. One way to achieve this scheme is by flipping the structure in a flip-chip fashion and exciting with a suitable photon energy that is strongly absorbed in the lower AlGaSb barrier layer before reaching the QW.

Table II. Fitting parameters from the rate equations $E_{QD}$, $E_b$ and $\Delta E_{QW-QD}$

| QD system | $E_{QD}$ (meV) | $E_b$ (meV) | $\Delta E_{QW-QD}$ (meV) |
|---|---|---|---|
| $QD_{5.7}$ | 114 | 5.0 | 93 |
| $QD_{6.1}$ | 86 | 4.3 | 100 |

### D. Theoretical description of the energy structure

In the following, we distinguish and analyze the 1D QW system and the 3D QDs separately. First, for the 1D QW, the bandgap undergoes a transition from an indirect to a direct state as the QW thickness ($t_{QW}$) increases. To examine this crossover, we initially constructed a one-dimensional model comprising a GaSb QW surrounded by AlGaSb. By calculating the energy of the CB for varying $t_{QW}$, we compare the energy levels of the Γ-band and the L-band to characterize the bandgap. When the $t_{QW}$ is small, the L-band resides below the Γ-band, indicating an indirect bandgap. However, as the $t_{QW}$ increases, the L-band surpasses the Γ-band, causing the latter to become the minimum of the CB, thus resulting in a direct bandgap. In this 1D system, the transition from an indirect to a direct bandgap occurs at approximately $t_{QW} = 5$ nm, aligning well with experimental observations.

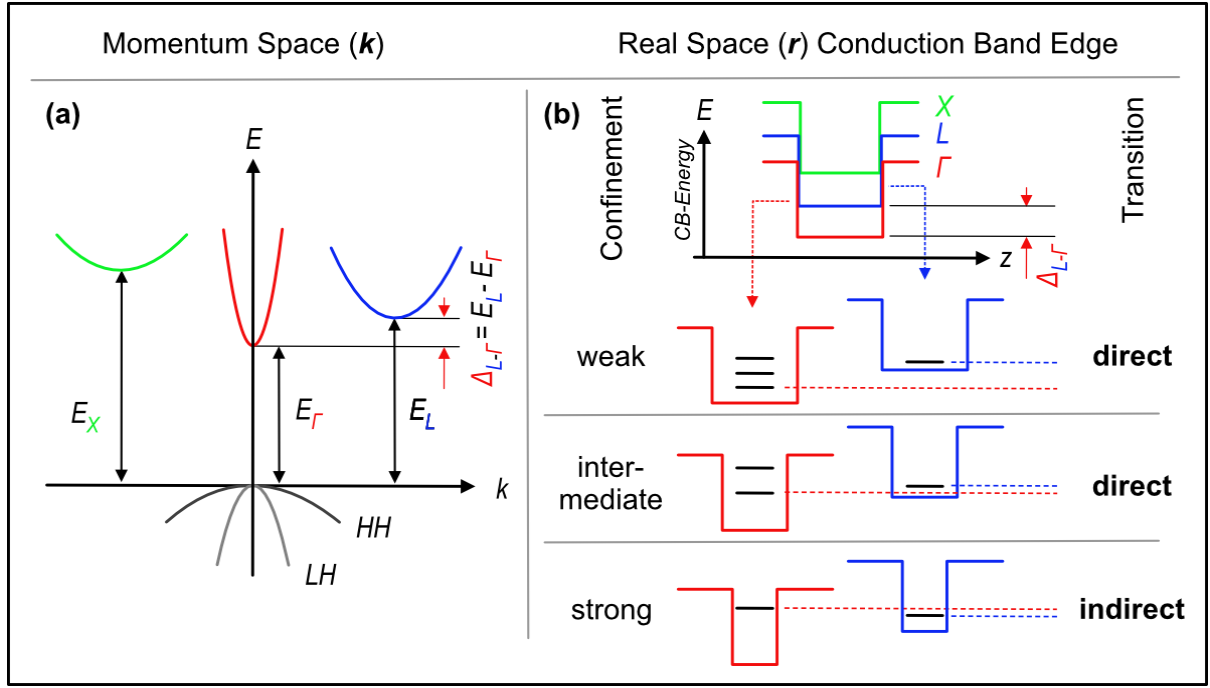

FIG. 10. (a) A simplified energy band diagram of GaSb bulk including heavy hole (HH) and light hole (LH) VBs and Γ, L, and X valleys in the CB. (b) Corresponding band-edge scheme in real space for three different QD sizes and related positions of confined electron Γ- and L-states: for large QDs the Γ-electron GS and first ES are below the L-GS. For the intermediate-size QDs still the Γ-electron GS lies below the L-GS electron. In both cases the lowest energy electron-hole optical transition is between the confined Γ-hole and Γ-electron ground states and, thus, direct. For the smallest QDs, a crossover between the Γ-GS and the L-GS occurs leading to an indirect transition as lowest possible optical transition with much smaller oscillator strength.

Subsequently, we proceeded to simulate a three-dimensional model by introducing a GaSb QD on top of the QW. After exploring various shapes available within the nextnano++ software, we selected a cone shape that closely resembles the experimental nanoholes. Similar to the QW case, we are now investigating the transition from direct to indirect optical transitions in the case of QDs. First of all, bulk GaSb represents a direct semiconductor, with the CB mininum at the Γ-point, the L-point energy 63 meV and the X-point energy 329 meV above the band-edge, respectively (Fig. 10(a)). However, in confined nanostructures, the positions of the Γ-GS and L-GS energies not only depend on the local Γ- or L-band-edges but also on the quantization energy. The quantization energy represents the energy difference between the respective band-edge and the corresponding GS energy. Since this quantization energy is roughly inversely proportional to the effective mass multiplied by the size of the nanostructure, we anticipate a transition from direct (Γ-hole to Γ-electron) to indirect (Γ-hole to L-electron) in narrower nanostructures. Fig. 10 illustrates this scenario for three different QD sizes (Fig. 10(b)), displaying the corresponding positions of the confined electron Γ- and L-states. In larger QDs, the Γ-electron GS and the first ES are both located below the L-GS. In moderate-sized QDs, the Γ-electron GS still remains below the L-GS. In both cases, the lowest energy electron-hole optical transition occurs between the confined Γ-hole and the Γ-electron ground states, resulting in a direct transition. However, in the smallest QDs, a crossover between the Γ-GS and the L-GS takes place, leading to an indirect transition as the lowest possible optical transition, which possesses significantly smaller oscillator strength.

Multiple systems were modeled with QDs of varying depths and diameters, and for each structure, we adjusted the $t_{QW}$ to identify the point at which the bandgap becomes direct. Fig. 11 illustrates how the crossover rapidly shifts with changes in QD depth ($t_{QD}$) and QD diameter ($d_{QD}$). On the left side of the graph, structures with a $t_{QD}$ = 10.5 nm exhibit a crossover occurring at a $t_{QW}$ close to 20 nm, which appears disproportionate. Thus, in this scenario, it can be concluded that the bandgap remains indirect for suitable QD and QW dimensions. Conversely, on the right side, for larger QDs, the crossover approaches zero swiftly, signifying a consistently direct bandgap. Only a limited number of structures in the middle, featuring depths near 12 nm, demonstrate more appropriate crossovers ranging between 4 and 8 nm.

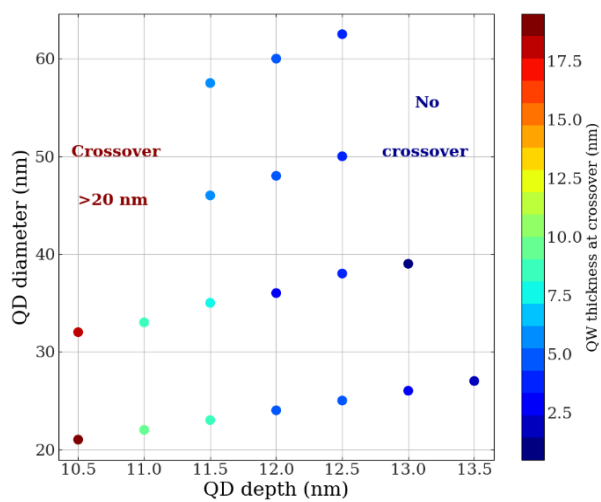

FIG. 11. QW thickness at the indirect-direct bandgap crossover depending on the QD depth and diameter.

To investigate the dipole transitions of our system, we examined systems with large QDs measuring more than 12 nm in diameter and thick QWs measuring more than 5 nm in thickness. These systems displayed a direct bandgap with QD emission. We began by conducting a parameter search and studied a total of 125 structures with varying $t_{QD}$ and $d_{QD}$, as well as different $t_{QW}$, to find the closest match to experimental results. We evaluated the dipole transitions between the first ten electron states and the first ten hole states, focusing on transitions thought to correspond to the experimentally-observed TDPL peaks: GS, ES and P.

In the following, we will focus on two series to exemplify the role of $t_{QW}$ for a fixed QD size (series A) and the role of QD size for fixed $t_{QW}$ (series B). Series A consists of a QD with $t_{QD}$ = 12 nm and $d_{QD}$ = 24 nm while the $t_{QW}$ ranges from 0 nm to 8 nm, thus covering the range of experimental values. Series B consists of a QW with 5.5 nm thickness and QD size ranging from 12 nm to 27 nm in depth and twice this number in diameter. In Figs. 12(a) and 12(b), we present the evolution of the single-particle energies for series A. As the $t_{QW}$ increases, there is a notable decline in $\Gamma$-electron energies coupled with an ascent in hole energies. This shift results in diminished recombination energies, evident in the computed absorption spectra depicted in Fig. 13. Additionally, the spacing between $\Gamma$-single-particle sub-levels narrows, causing the primary spectral features to converge more closely.

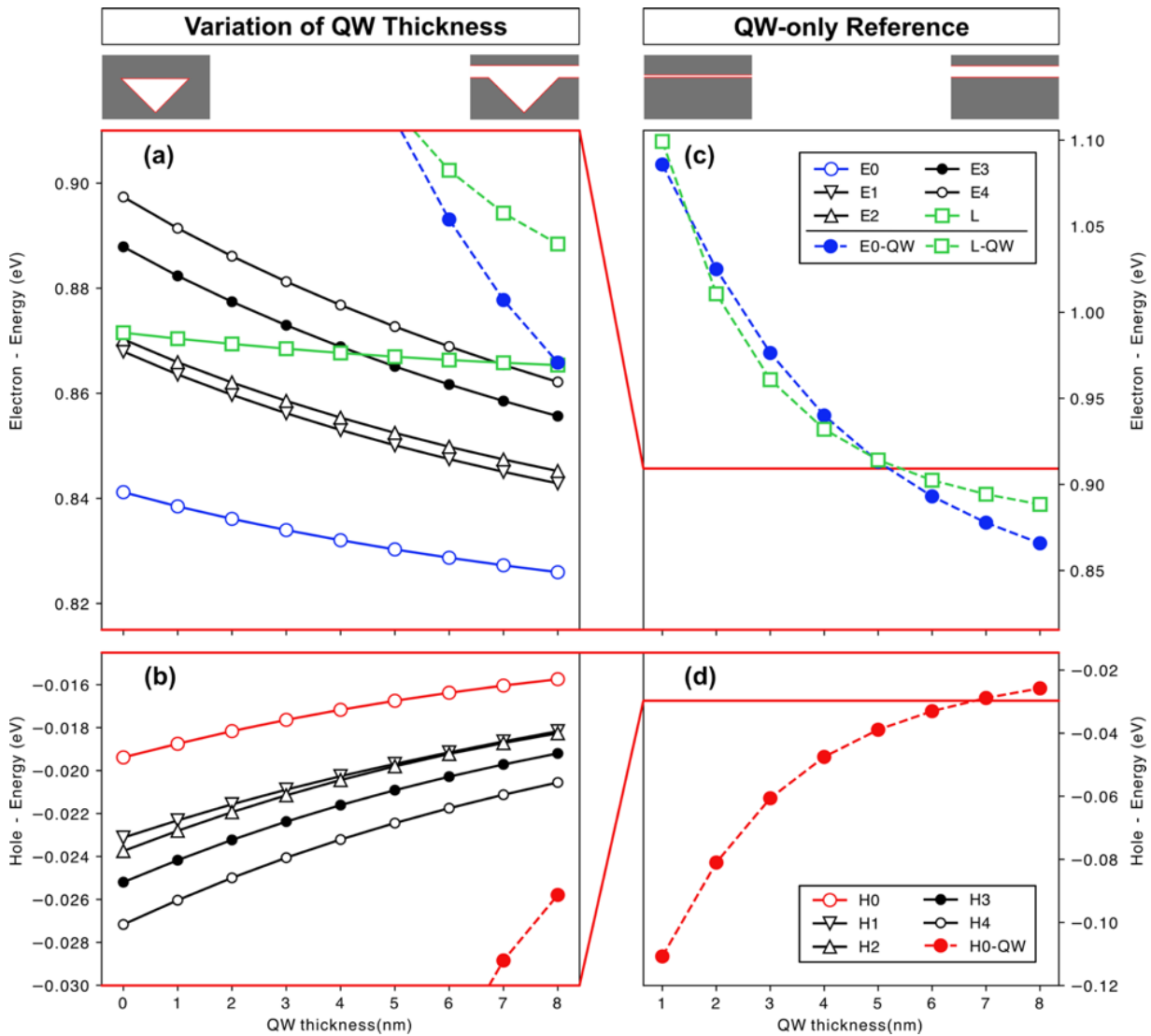

Fig. 12. (a) $\Gamma$, and L-electron- and (b) $\Gamma$-hole energies of Series A as function of QW thickness. (c,d) energies of the $\Gamma$-electron and -hole ground states, along with the lowest L-electron state for the unaltered QWs, plotted against QW thickness.

The energies of the L-electron GS, in contrast, display a more gradual decline as the $t_{QW}$ increases. This can be attributed to its smaller effective mass and, consequently, a reduced sensitivity to confinement effects. As a result, in emission spectra where Γ-electron states above the L-states don't contribute to the emission, the potential number of QD-emission peaks diminishes with a reduction in $t_{QW}$. For reference, Figs. 12(c) and 12(d) illustrate the energies of the Γ-electron and -hole ground states, along with the lowest L-electron state for the unaltered QWs plotted against $t_{QW}$.

The role of QD size (series B) with respect to the electron and hole energies is shown in Fig. 15, where the results for Γ-electron and -hole states together with the fundamental L-electron are plotted against the QD depth. A pronounced quantum-size effect is observed, especially for the Γ-electron states. In contrast, this effect is less distinct for the holes and in particular the L-electron states. Consequently, the relative energetic position of the L-electron state varies in relation to the Γ-electron state throughout this series. Notably, at a depth of 7 nm (which is not depicted here), the L-state descends below the most fundamental Γ-electron state. At this juncture, we would not anticipate significant light emission from the QD-confined Γ-states. This is consistent with our experimental findings of a critical QD dimensions below which no PL is observed [14].

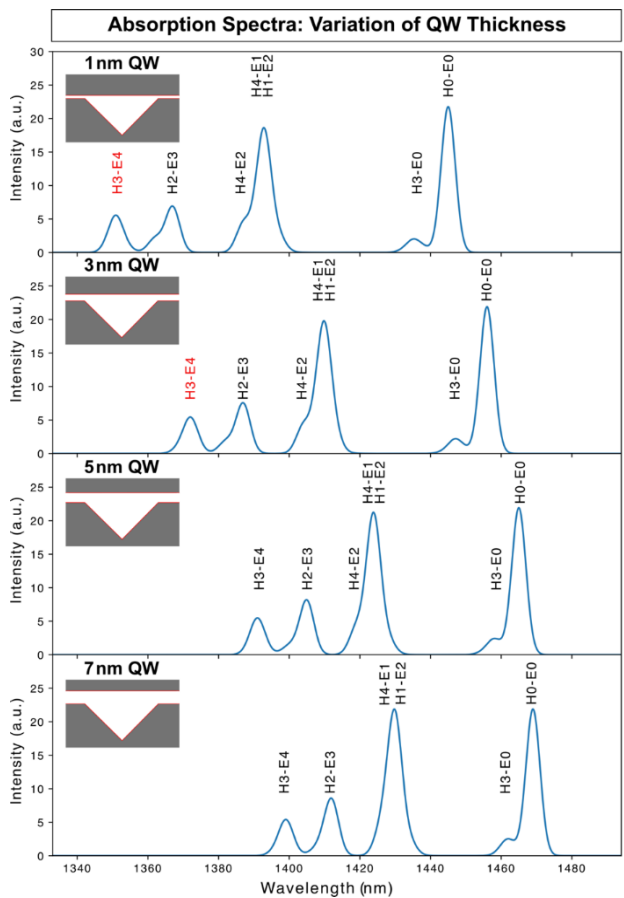

Fig. 13. The calculated absorption spectra for parts of Series A are presented with labels indicating the primary Γ-electron-hole transitions. Transitions labeled in red are less likely to appear in emission spectra since the associated Γ-electron state resides above the most fundamental L-electron state.

The absorption spectra displayed in Fig. 13 closely mirrors the observed PL spectra under high excitation density. This allows us to predominantly attribute the GS peak to the H0-E0 transition, and the ES peak to the transitions H1-E2, H2-E2 and H4-E1. Intriguingly, no major spectral features are discernible within the energy range of the observed P peak, which would primarily involve E4 or E5 state.

The QD that aligns best with our spectroscopic observations has a $t_{QD}$ = 12 nm, $d_{QD}$ = 24 nm, and $t_{QW}$ ranging from 5 nm to 6 nm. This is consistent with the available experimental TDPL data (Fig. 6).

In Fig. 14, we display the probability densities of the wavefunctions for five Γ-electron and five Γ-hole states, observed from various crystallographic perspectives. The Γ-electron states are easily categorized by their number of nodal planes: s-like with zero nodal planes (E0), p-like with one nodal plane (E1, E2), and d-like with two nodal planes (E3, E4). This straightforward classification is not feasible for the showcased hole states, with the exception of H0, which exhibits s-like characteristics. Due to the interplay between heavy-hole (HH) and light-hole (LH) properties, a hole state often amalgamates both HH and LH contributions, usually blending HH and LH percentages with different symmetries. As a consequence, there are no clear selection rules for the absorption spectra. An electron state with a particular symmetry often aligns with a symmetry-compatible component in either the HH or LH portion of a specific hole state. As a result, in Fig. 13 and Fig. 16, we have only labeled the predominant contributions to the absorption spectra.

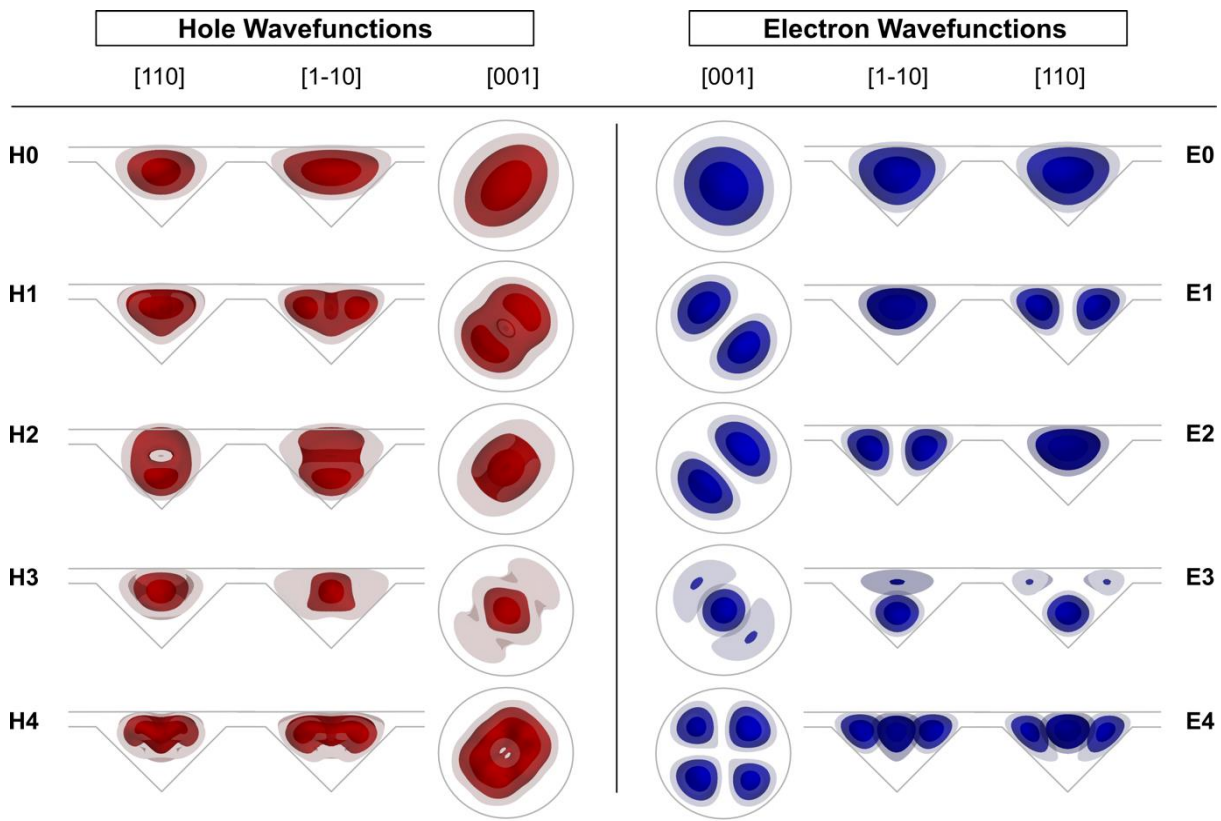

Fig. 14. Γ-hole (red) and electron (blue) wavefunctions (probability densities) for a QD with a depth of 12 nm and a diameter of 24 nm, situated beneath a 5 nm thick QW. The viewing axes are [110], [1-10], and [001].

Our research determines that as the overall dimensions of the system – including $t_{QD}$, $d_{QD}$ and $t_{QW}$ – expand, the energy of the ground state diminishes. As depicted in Fig. 15, with varying QD diameters, an uptick in QD depth corresponds to a drop in the energy of the GS peak. A similar pattern emerges across different QW thicknesses, distinguished only by a minor decline in the E0-H0 transition energy as the QW thickens. Peak shifts are also noticeable (Fig. 13). For structures with a 12 nm depth, an increased diameter causes the ES peak to drift closer to GS.

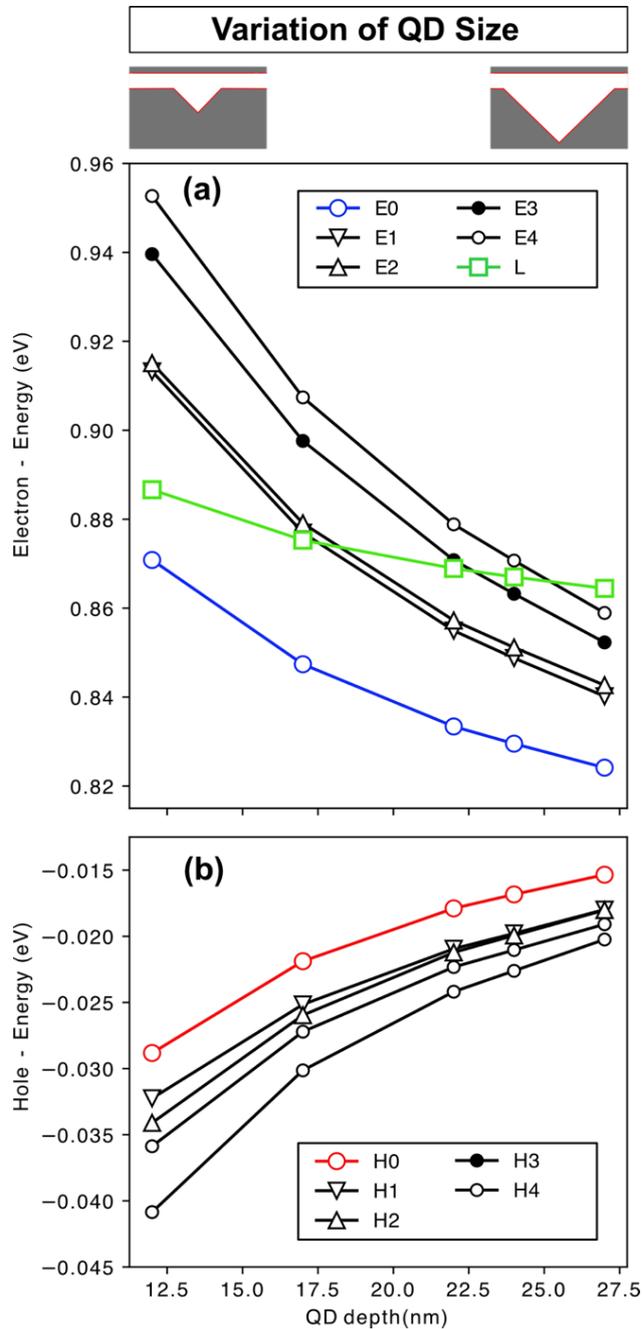

Fig. 15. (a) Evolution of Γ- and L-electron- and (b) hole states energies of Series B as function of QD-depth.

Lastly, we homed in on simulating structures that mirrored the dimensions of the epitaxially-grown nanostructures used in the PL experiments. We simulated various QDs with depths ranging from 12 nm to 27 nm, maintaining a consistent diameter-to-depth ratio of 2. We then gauged the transition energy and oscillator strength of the E0-H0 transition across different QW thicknesses. The E0-H0 transition energy fluctuates between 0.899 eV and 0.839 eV, which equates to wavelengths between 1378 nm and 1477 nm.

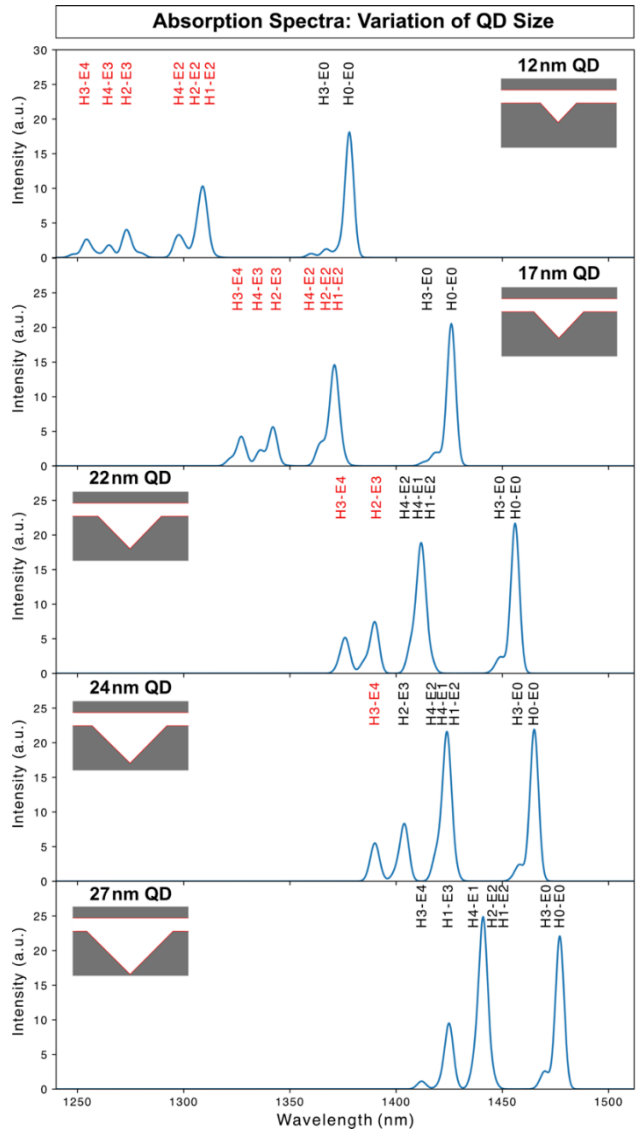

Fig. 16. The calculated absorption spectra for parts of Series B are presented with labels indicating the primary Γ-electron-hole transitions. Transitions labeled in red are less likely to appear in emission spectra since the associated Γ-electron state resides above the most fundamental L-electron state.

## IV. SUMMARY AND CONCLUSION

A detailed description of the electronic structure of GaSb QDs using a rate equation model supported by experimental temperature-dependent PL data was introduced. The presence of a considerable energy barrier for non-resonantly generated carriers to move into QD states points towards the need for resonant excitation schemes to efficiently generate single photons. Use of either coherent resonant excitation or incoherent quasi-resonant excitation should provide true single exciton lifetimes, with the former being the most efficient excitation-decay scheme and thus preferential for generating optimal single- and entangled-photon metrics. Further, we applied the eight-band k·p model to calculate energy levels and wavefunctions of bound electron and hole states in the investigate heterostructures, with an emphasis on various shapes and material compositions. Utilizing the nextnano++ software, quantum nanostructures were simulated, focusing on the transitions between direct and indirect bandgaps in 1D QWs and 3D QDs. The research reveals that as QD and QW dimensions change, there are significant shifts in energy states and optical properties. The best-matching model to experimental results features a QD with a depth of 12 nm, a diameter of 24 nm, and a QW thickness between 5 nm and 6 nm. The experimental results obtained here will prove to be useful when designing QDs that emit at longer wavelengths, for example towards 2 μm. Additionally, new spin-qubit interfaces based on single GaSb QDs will benefit from detailed information about the electronic structure of this system.

## ACKNOWLEDGEMENTS

AC, JH, MG, and TH acknowledge financial support from the Academy of Finland project QuantSi (Decision No. 323989) and Business Finland co-innovation project QuTI. LL and AS acknowledge the MSCA-ITN-2020 Funding Scheme from the European Union's Horizon 2020 research and innovation programme under grant agreement ID: 956548 (Quantimony). nextnano GmbH, in particular Stefan Birner and Takuma Sato, are acknowledged for providing their software (nextnano++) for parts of this work and for their kind support.